\documentclass[namedreferences]{solarphysics}

\usepackage[hyperref,optionalrh]{spr-sola-addons} 
\usepackage{graphicx}        
\usepackage{color}           

\newcommand{\aap}{    {\it Astron. Astrophys.}}

\newcommand{\apj}{    {\it Astrophys. J.}}

\newcommand{\grl}{    {\it Geophys. Res. Lett.}}

\newcommand{\jgr}{    {\it J. Geophys. Res.}}

\newcommand{\solphys}{{\it Solar Phys.}}
 
\newcommand{\ssr}{    {\it Space Sci. Rev.}} 

\begin{document}

\begin{article}

\begin{opening}

\title{Real-time solar wind prediction based on SDO/AIA coronal hole data\\
 {\it Solar Physics}}

\author{T.~\surname{Rotter}$^{1}$\sep
        A.M.~\surname{Veronig}$^{1}$\sep
        M.~\surname{Temmer}$^{1}$\sep
        B.~\surname{Vr{\v s}nak$^{2}$\sep}   
       }
\runningauthor{T. Rotter et al.}
\runningtitle{Solar Wind Predictions}

   \institute{$^{1}$ Kanzelh\"ohe Observatory-IGAM, Institute of Physics, University of Graz, Universit\"atsplatz 5, 8010 Graz, Austria\\    email:		\url{thomas.rotter@uni-graz.at}, email: \url{asv@igam.uni-graz.at}\\ 
              $^{2}$ Hvar Observatory, Faculty of Geodesy, Zagreb, Croatia, email: \url{bvrsnak@geodet.geof.hr}}

\begin{abstract}
We present an empirical model based on the visible area covered by coronal holes close to the central meridian  in order to predict the solar wind speed at 1 AU with a lead time up to four days in advance with a 1-h time resolution.
Linear prediction functions are used to relate coronal hole areas to solar wind speed. The function parameters are automatically adapted by using the information from the previous 3 Carrington Rotations. Thus the algorithm automatically reacts on the changes of the solar wind speed during different phases of the solar cycle.  The adaptive algorithm has been applied to and tested on SDO/AIA-193 \AA \, observations and ACE measurements during the years $2011-2013$,  covering 41 Carrington Rotations.  The solar wind speed arrival time is delayed and needs on average $4.02\pm0.5$ days to reach Earth.  The algorithm produces good predictions for the 156 solar wind high speed streams peak amplitudes with correlation coefficients of $cc\approx0.60$. For 80$\%$ of the peaks, the predicted arrival matches within a time window of 0.5 days of the ACE \textit{in situ} measurements. The same algorithm, using linear predictions, was also applied to predict the magnetic field strength from coronal hole areas but did not give reliable predictions ($cc\approx0.2$).

\end{abstract}
\end{opening}

\section{Introduction}      \label{Introduction} 
The Sun's corona is constantly emanating magnetized plasma into space, the solar wind, consisting of charged particles (protons, electrons, Helium ions). The slow solar wind has typical speeds in the range $300$-$400$~km s$^{-1}$, whereas the so-called high-speed solar wind streams (HSSs) may reach speeds up to $800$~km s$^{-1}$. HSSs originate from coronal holes (CHs) on the Sun (\opencite{Krieger1973}). Since the $1970$s, CHs are known for being long-lived and rigidly rotating phenomena in the solar corona (\opencite{Krieger1973,Neupert1974,Nolte1976}), magnetically ``open'' to interplanetary space, allowing the solar wind particles to escape from the Sun (\opencite{Tsurutani1995,Gosling1999,Cranmer2009}). The plasma temperature and density in CHs is low compared to their surroundings, and thus they appear as dark areas in the corona (\opencite{Munro1972,Cranmer2009}).
They can be observed in space-based X-ray (\opencite{Wilcox1968,Altschuler1972,Hundhausen1972}) and EUV images (\opencite{Newkirk1967,Tousey1968,DelZanna1999}).  More recent studies on the outflow dynamics and Doppler shifts of coronal holes improved the understanding of CHs being the source of the fast solar wind (\opencite{Hassler1999,peter1999,xia2003,aiouaz2005}). 

Studying coronal holes and their associated HSSs is an important task, since in combination with the Sun's rotation they shape the solar wind distribution in interplanetary space and are the dominant contributors to space weather disturbances at times of quiet solar activity, due to recurrent geomagnetic storm activity. The distribution of the solar wind is also decisive for deriving the propagation behavior of CMEs (\opencite{Temmer2011}).
The relationship between the size and location of CHs in the corona and the solar wind parameters and geomagnetic effects measured at 1 AU has been investigated by several groups (\opencite{Nolte1976,Robbins2006,Vrsnak2007a,Vrsnak2007b,Luo2008,Abramenko2009,Obridko2009,deToma2011,Verbanac2011,lowder2014}). In the early phases of coronal holes identification, CHs were visually tracked by experienced observers (\opencite{harvey2002,McIntosh2003}). Recently, several groups tried to automate the process for the identification and detection of coronal holes using different approaches such as perimeter tracing (\opencite{Kirk2009}), fuzzy clustering (\opencite{Barra2009}), multichannel segmentation (\opencite{delouille2007}), edge-based segmentation (\opencite{Scholl2008}), intensity thresholding (\opencite{Krista2009,deToma2011,Rotter2012}) and magnetic track-boundaries (\opencite{lowder2014}). \\
One way to model the physical parameters of the solar wind are MHD models of the corona and heliosphere, using synoptic solar magnetic field maps as input,  such as ENLIL (\opencite{Odstrcil2009}). Since large and long-lived CH structure cause clear signatures in the solar wind parameters measured at 1 AU they can be used to derive empirical relationships between the CH area on the Sun and the solar wind parameters. To this aim, we extract the areas of coronal holes from remote sensing EUV imagery and relate these results to \textit{in situ} measurements of the solar wind parameters. We developed an adaptive algorithm, using linear relations that react on the solar cycle variations. In the present paper we evaluate and adjust the algorithm based on observations during the beginning of solar cycle 24, the years 2011-2013. In Section 2 we describe the data used for our study. Section 3 describes the adaptive linear prediction method. The method is applied to predictions of solar wind speed and magnetic field strength at 1 AU. As will be shown, reliable predictions are obtained for $v$ and not for $B$. The results of the predictions are shown in Section 4. Conclusions are drawn and discussed in Section 5.

\section{Data} \label{Data}
Our analysis is based on the following data sets: The fractional coronal hole areas, $A$ (the area of coronal holes calculated as a fraction over a slice of $\pm7.5^{\circ}$), derived from EUV images and the solar wind speed $v$ and magnetic field strength $B$ measured \textit{in situ} at L1. As described in \inlinecite{Rotter2012}, we adjusted the coronal hole extraction algorithm to process the $1$k by $1$k \textit{Solar Dynamic Observatory} SDO \textit{Atmospheric Imaging Assembly} AIA-193\AA \, (\inlinecite{Lemen2012}) quicklook images, accessible in near real-time at \textit{Royal Observatory of Belgium} ROB (\url{http://sdoatsidc.oma.be/}), or at the \textit{Joint Science Operations Center} JSOC (\url{http://jsoc.stanford.edu/}).

For our purposes we work with images with an approximate cadence of 1~h using the SDO software routines provided within \texttt{SolarSoft}. We utilize the solar wind data measured \textit{in situ} at 1 AU by Solar Wind Electron Proton and Alpha Monitor (SWEPAM; \opencite{SWEPAM1998}) and the magnetometer instrument (MAG; \opencite{MAG1998}) on board the {\it Advanced Composition Explorer} (ACE; \opencite{ACE1998}). The hourly-averaged level-2 ACE data, available at \url{http://www.srl.caltech.edu/ACE/ASC/level2/}, with a resolution of 1~h is used. We analyse the data available 2010 October 3 to 2013 December 31.  In total, we processed a set of $28512$ AIA-193 \AA \, images.

\section{Methods} \label{Methods}
The main goal is to fully automate the process of identification and extraction of coronal hole regions in solar EUV images and to derive near real-time predictions for solar wind parameters at 1 AU up to 4 days in advance. The algorithm performs the following main steps:

\begin{enumerate}
        \item Ftp-download and image selection.
        \item Process selected images and derive fractional CH areas.
        \item Derive prediction parameters via adaptive algorithm.
\end{enumerate}

Step 1: Automatically download and select images for our purpose. In case of the SDO/AIA 193\AA \, quicklook images, we download images in quasi real-time with a cadence of 1~hour. The quicklook images are available with a delay of  $\sim$3 hours after the observation. The delay is caused by data downlink and data preparation. Images are checked for their quality. Data including bad pixels, flare emission, or eclipse data are excluded from further analysis.

\begin{figure}[ht]    
					\centering
					\includegraphics[width=\linewidth]{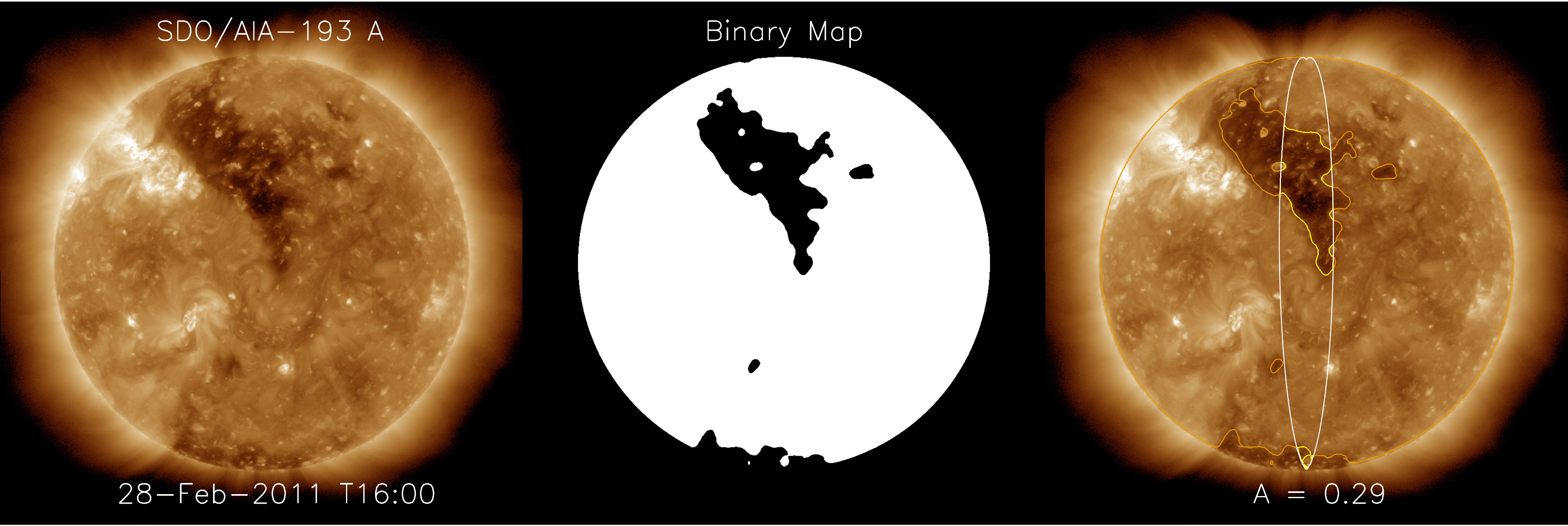}
						\caption{Left: Calibrated SDO/AIA-193 \AA \, image on 28-02-2011. Middle: Resulting binary map using histogram thresholding. Right: CH binary map overlayed on the original AIA image. The yellow lines mark the derived fractional area $A$ in the considered meridional slice [$\pm7.5^{\circ}$]. }
				\label{fig001}
   \end{figure}

\begin{figure}[ht]    
					\centering
					\includegraphics[width=\linewidth]{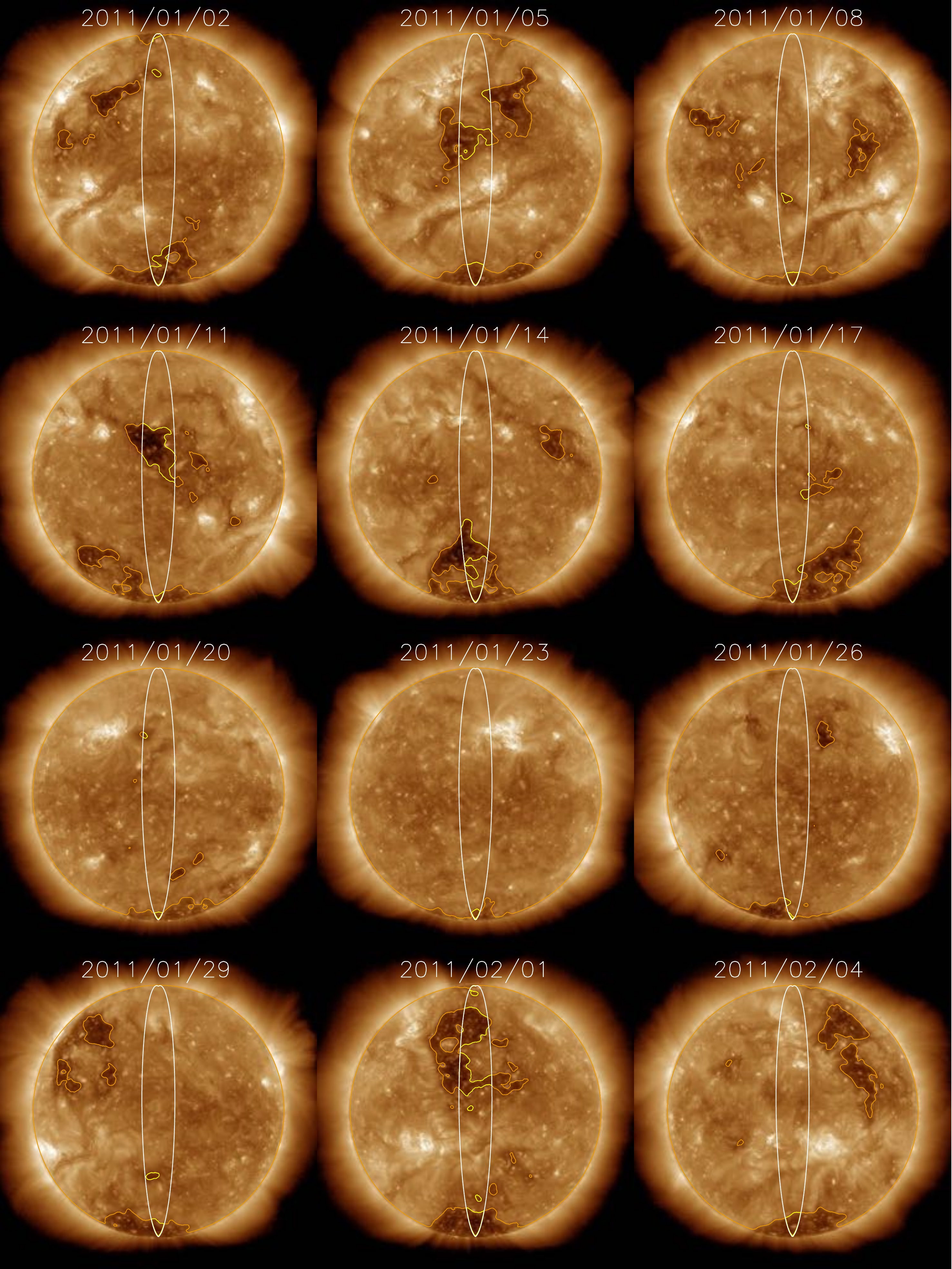}
						\caption{Examples of CH detections for SDO/AIA-193 \AA\,.}
				\label{fig002}
   \end{figure}

Step 2: 
The CH extraction algorithm utilizes a histogram-based thresholding segmentation technique to detect the boundaries of CHs (\inlinecite{Rotter2012}). Figure~\ref{fig001} illustrates the major steps in image processing. On the left panel a fully calibrated SDO/AIA-193\AA \, taken on 28 February 2011 is shown. The algorithm searches for a dominantlocal minimum  in the intensity distribution for the EUV image.  On average it lies between $55\pm15$ digital numbers (DN), consistent with the findings of \inlinecite{lowder2014}. If the algorithm is not able to detect a well defined minimum in the distribution, which is the case for images where $\lesssim$~5$\%$ of the whole solar disc  is covered by coronal hole pixels, the threshold is set to 55 DNs. In approximatly 18$\%$ of all processed images, the algorithm was set to the fix threshold.
The resulting binary map undergoes mathematical morphological image processing techniques to remove small anomalies (\texttt{Erosion}) and to gradually fill small holes in-between detected CH regions (\texttt{Dilation}), which is illustrated in the middle panel. The right panel shows the original SDO/AIA image overlayed with the processed CH binary map. The fractional CH area, $A$, is derived inside the meridional slice $\pm7.5^{\circ}$ for the EUV image corresponding to the solar rotation over $\sim$1 day. All identified CH pixels inside the given slice are summed up and divided by the total number of pixels in the slice (see also \inlinecite{Rotter2012}). Figure~\ref{fig002} illustrates examples of the image processing algorithm on SDO/AIA-193\AA\, for the beginning of the year 2011. 

Step 3:
The forecast is based on deriving the delay between the fractional CH area time series $A_{(t)}$ and the solar wind speed $v_{(t)}$ \textit{in situ} measured by ACE using cross-correlation. Linear least-square fits tend to underestimate and smooth out peak values, and are thus not well suited as a functional form for the forecasting procedure of peaks in the solar wind speed. To predict the solar wind speed, we shift the CH area time series with the derived lag $\tau$ and apply a linear function $v_{(t + \tau)} = c_{0} + c_{1}A_{(t)}$ (cf. \opencite{Vrsnak2007a}). The parameters $c_{0}$ and $c_{1}$ are obtained by searching for the minimum and maximum values of the solar wind speeds and the corresponding CH fractional areas in preceding periods. We exclude values of high solar wind speeds ($>800$~km s$^{-1}$) that are connected to CH areas $A<0.1$, as they most probably arise due to interplanetary CMEs and not due to HSSs. \\
The algorithm reacts on variations in the relation between the CHs and solar wind speed.
To account for the variations over the Solar Cycle, the algorithm  is automatically adapted by using the information from the three preceding Carrington Rotations (CRs) to update the prediction parameters $c_{0}$ and $c_{1}$ and the lag $\tau$ used for the predictions of the solar wind speed. These parameters are automatically updated by using a moving window of the duration of three CRs, sliding in steps of one CR. For the prediction function of a subsequent CR, we basically use the CH area, $A_{(t)}$ and solar wind speed $v_{(t)}$ space covered by the three preceding CRs. Testing showed that the adaptive algorithm performs best when using the three CRs moving window. It covers a long enough time frame to catch solar wind speed variations caused by CHs.

\begin{figure}[ht]    
						\includegraphics[width = 0.82\linewidth, angle = 90]{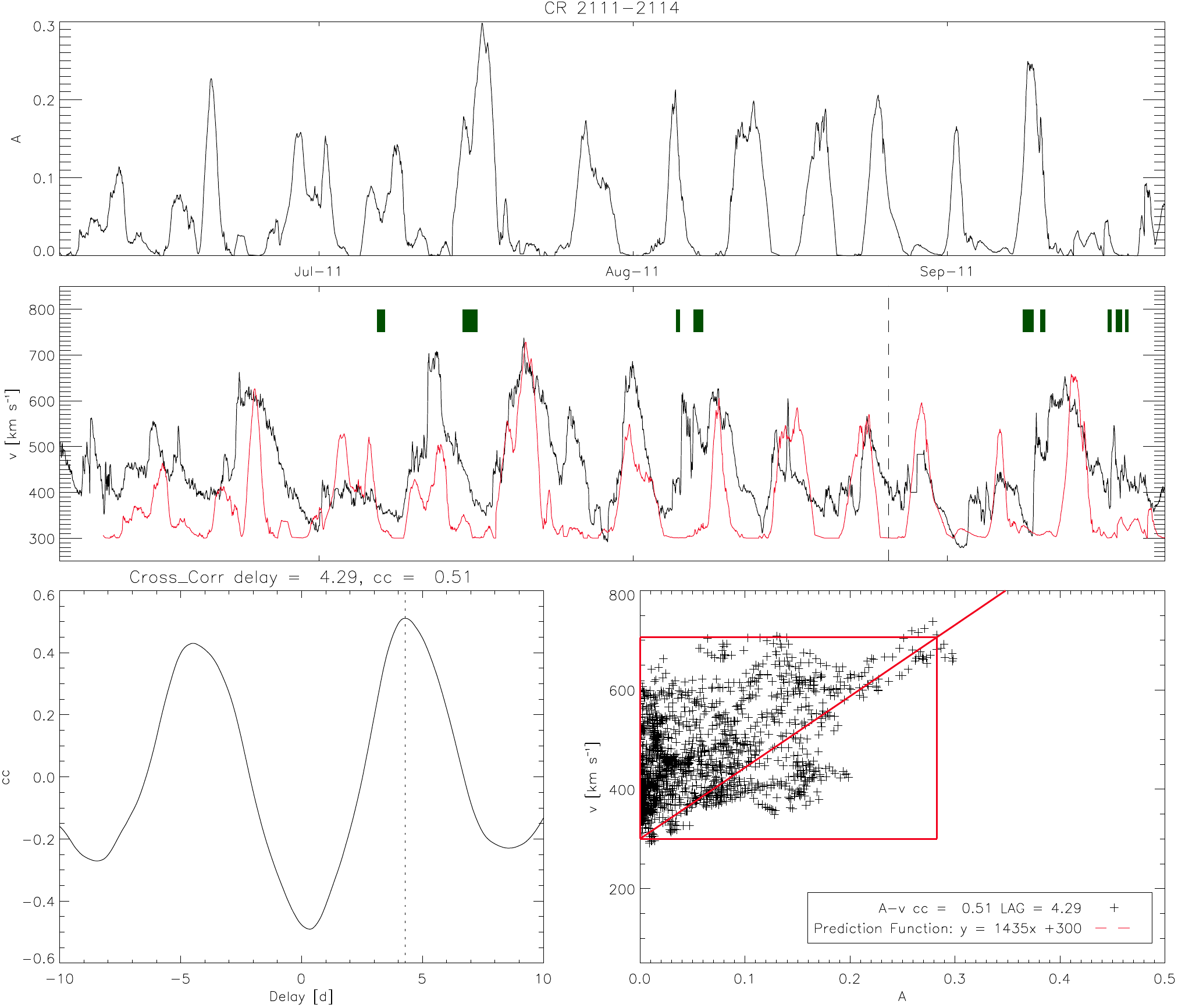}
		      	\caption{Top: CH fractional areas. Middle: Solar wind speed measured (black line) and predictions (red line) at 1 AU for the time range CR2111-2114. The dashed vertical line shows the boundary when the predictions for the following CR starts. The green bars indicate identified ICMEs at 1 AU (from Richardson and Cane list). Bottom left: the cross-correlation coefficient and the derived time lag in days. Bottom right: Scatter plot of the time lagged fractional CH areas and the measured solar wind speed. The prediction function (red line) and its linear function coefficients given as inset. }
						\label{fig003}
   \end{figure}

\begin{figure}[ht]    
						\includegraphics[width = 0.82\linewidth, angle = 90]{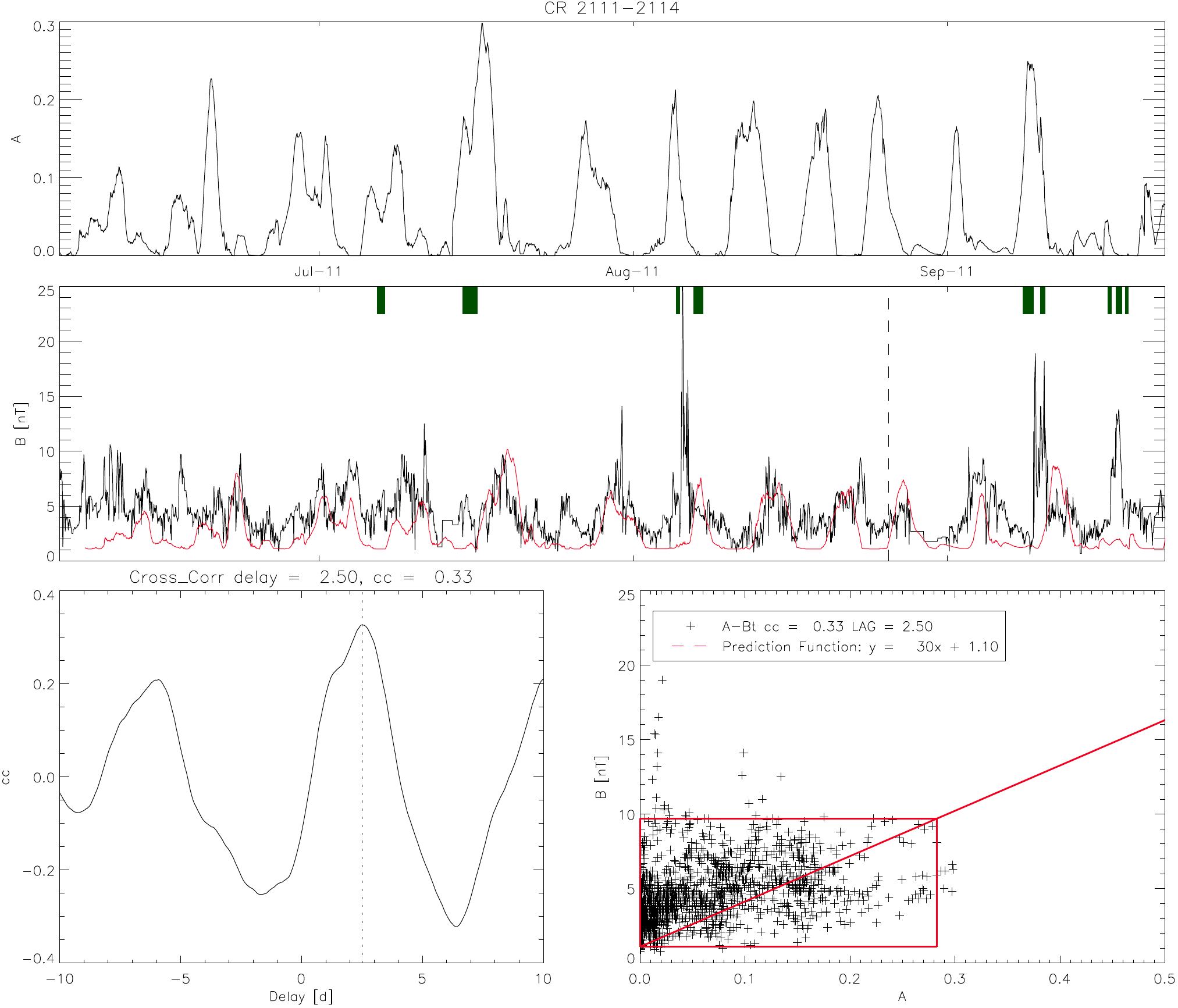}
	      	\caption{Same as Fig ~\ref{fig003} for the magnetic field strength.}
				\label{fig004}
   \end{figure}

The procedure is illustrated in Figure~\ref{fig003} for the solar wind speed. The dashed vertical line indicates the beginning of the predictions for the CR~$2114$ based on the preceding CRs $2111$-$2113$. The cross-correlation of the fractional CH areas $A_{(t)}$ (top panel) and \textit{in situ} measurements by ACE (middle panel) delivers the time delay (bottom left). We shift the $A_{(t)}$ time-series with the corresponding time lag $\tau$ by $4.29$ days. Similar to figure~\ref{fig003}, we applied the adaptive algorithm for the same time (CRs $2111$-$2113$) on the magnetic field strength, which can be seen in Figure~\ref{fig004}. For $B$ we excluded values $> 15nT$ that are connected to CH areas $A<0.1$.  The red rectangle covers the boundaries used for the prediction functions (bottom right). We exclude the $\sim1\%$ of extreme values. The green bars indicate identified ICMEs at 1 AU (\inlinecite{richardson2003}; \url{http://www.srl.caltech.edu/ACE/ASC/DATA/level3/icmetable2.htm}).

\begin{figure}[ht]    
						\includegraphics[width = \linewidth]{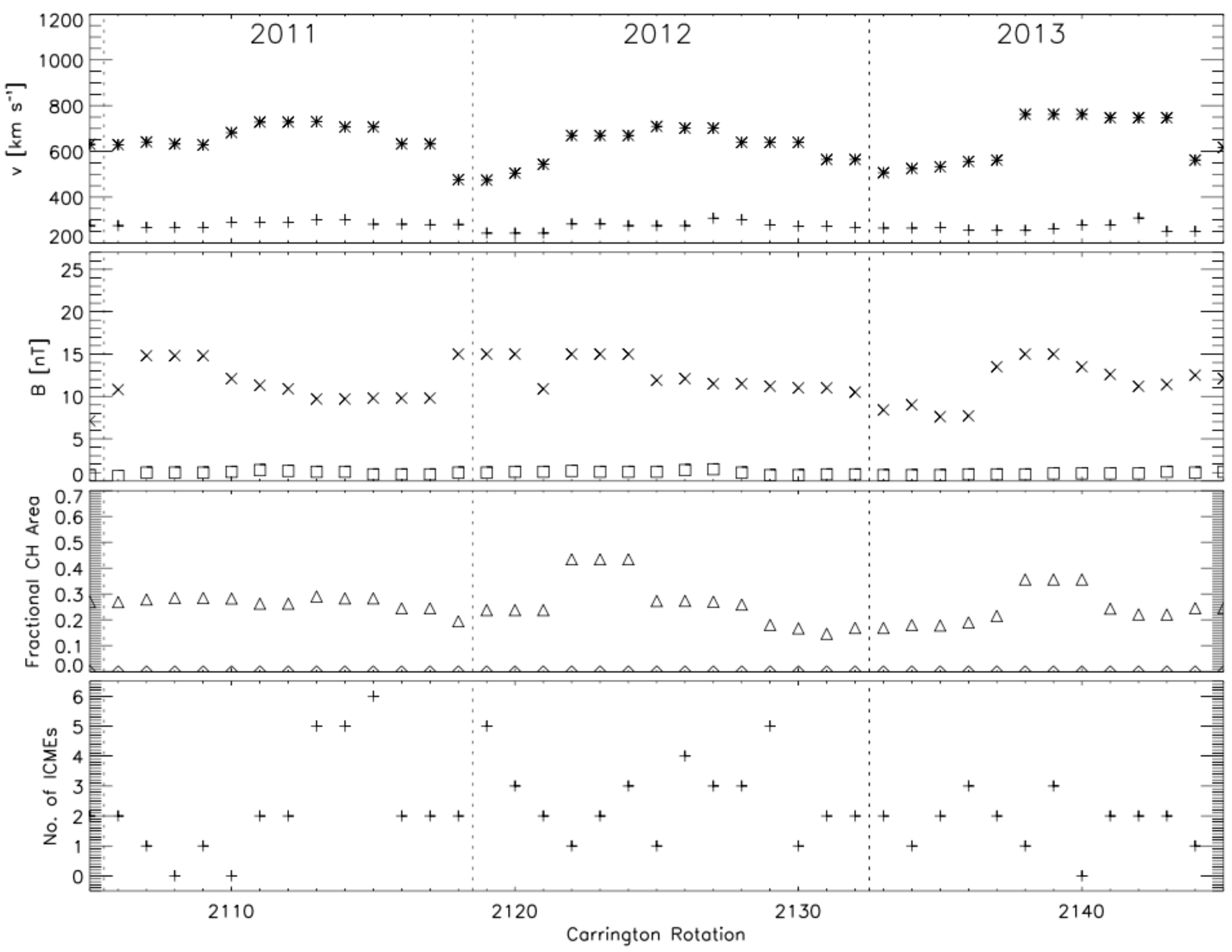}
					  	\caption{Minimum and maximum values of the prediction function (see Figures \ref{fig003} and \ref{fig004}) of the solar wind speed measurements (top), total magnetic field strength (middle) and the corresponding fractional CH areas in the sliding windows covering 41 consecutive~CRs. Number of ICMEs per CR (lowest panel).}
				\label{fig005}
   \end{figure}

Figure~\ref{fig005} shows the evolution of the parameters derived  for the prediction functions for each Carrington Rotation. Applying the adaptive algorithm to 41 CRs we found an average time lag of $4.07\pm0.07$ days for $v$ and $1.98\pm0.03$ for $B$. We used the knowledge of the average delays and set time lag boundaries for $v$, between $3.5-5$ days and for $B$ between $1.5-2.5$ days. If the delay derived from the previous 3 CRs does not lie between the given boundaries the delay is set to the average value of 4 days in the case of the solar wind speed and 2 days for the magnetic field strength. This only occurred twice for the solar wind speed and 7 times for the magnetic field parameter.

\section{Results} \label{Results}
 \begin{figure}[ht]  
					\includegraphics[width=\linewidth]{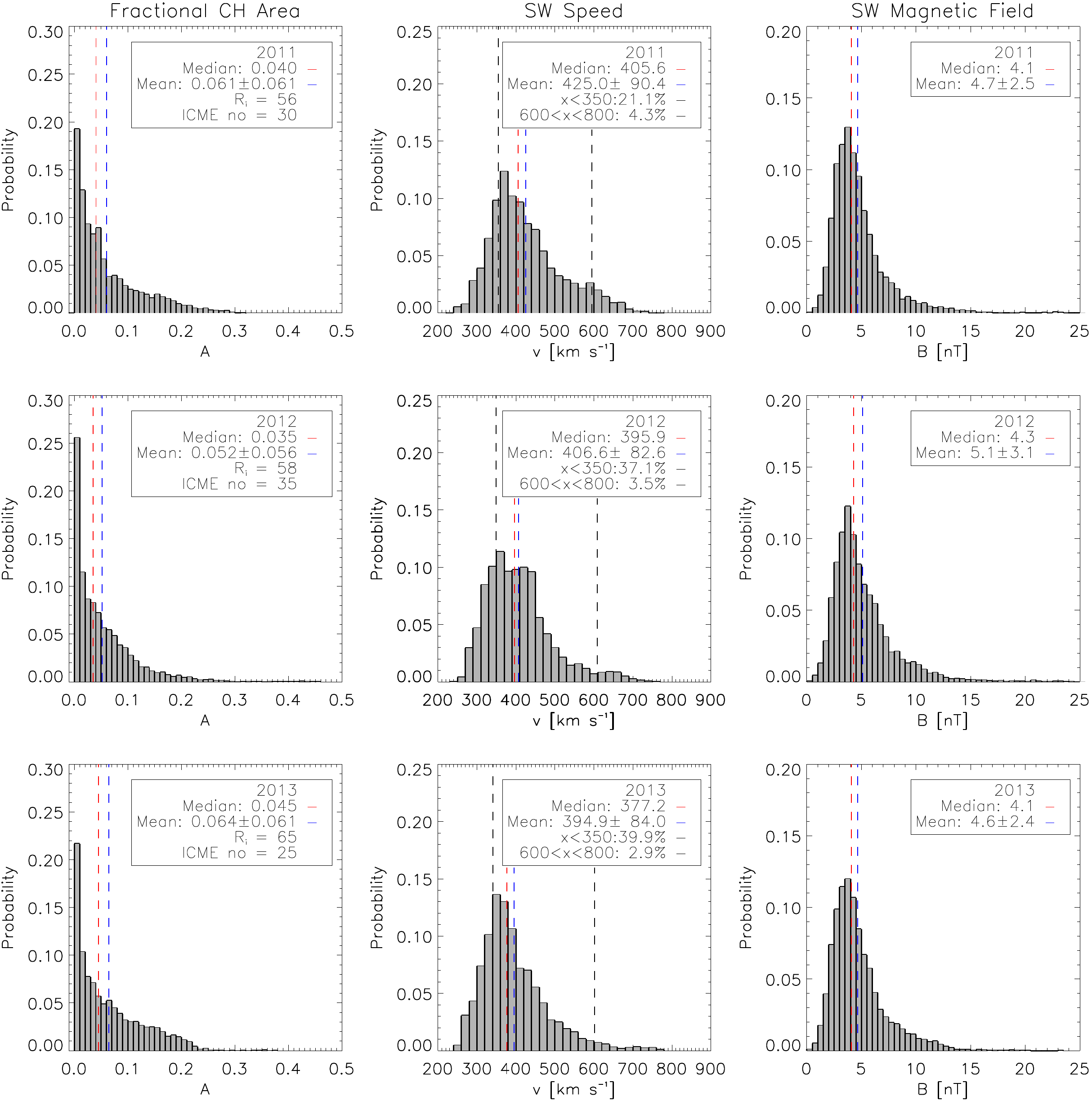}
      	\caption{From left to right: Probability distributions of fractional CH areas $A$, solar wind speed $v$ and total magnetic field strength $B$ for the years 2011 (top), 2012 (middle), 2013 (bottom) from SDO/AIA-193\AA \,, and ACE, respectively arithmetic mean and median values are indicated as blue and red lines. The yearly mean total sunspot numbers R$_{i}$ (SILSO), and number of identified ICMEs at 1 AU (from Richardson and Cane list)  as given in the insets in row 1.}
				\label{fig006}
   \end{figure}
	
In Figure \ref{fig006} we show the distribution of the fractional CH areas (left columns), the \textit{in situ} measurements by ACE for the solar wind speed (middle columns) and magnetic field strength (right columns) for the years $2011$-$2013$. The yearly mean total sunspot numbers, $R_{i}$ taken from the SILSO (\url{http://www.sidc.be/silso/}), rising from $R_{i} = 56$ up to $R_{i} = 65$ over the years $2011$-$2013$, indicating the slow rise of solar activity of cycle no.~$24$. The average size of the fractional coronal hole area $A$ and the magnetic field strength vary slightly over the years. The average solar wind speed decreases from $v = 424$~km s$^{-1}$ to $v = 395$~km s$^{-1}$. We also find a change in the occurrence rate of high solar wind speeds ($600-800$~km s$^{-1}$) which is $4.4\%$ in the year $2011$ down to $2.9\%$ in the year 2013. We measured an increase of low solar wind speeds ($<350$~km s$^{-1}$) over the years, increasing from  $21\%$ in 2011 to  $40\%$ in 2013. 

\begin{figure}[ht]   
					\includegraphics[width =0.8\linewidth, angle = 90]{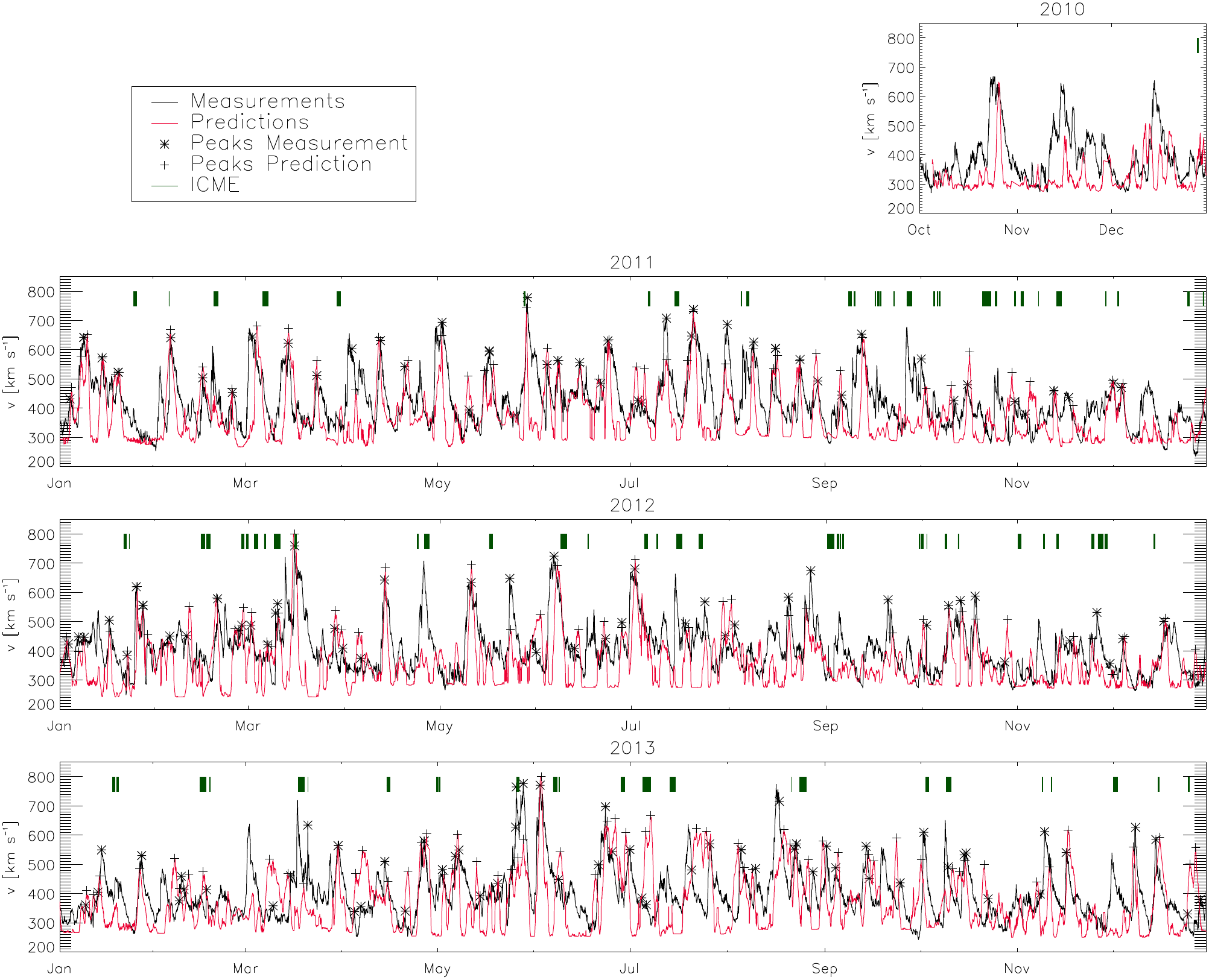}
      	\caption{Solar wind speed $v$ measured by ACE (black lines) and predictions derived from CH areas detected in SDO/AIA 193 \AA \, images (red lines). Cross-signs indicate peaks in the predicted solar wind speed and stars indicate peaks \textit{in situ} measured solar wind speed. Start and end time of identified ICMEs at 1 AU from Richardson/Cane list (green bars).}
				\label{fig007}
   \end{figure}

\begin{figure}[ht]  
					\includegraphics[width =0.8\linewidth, angle = 90]{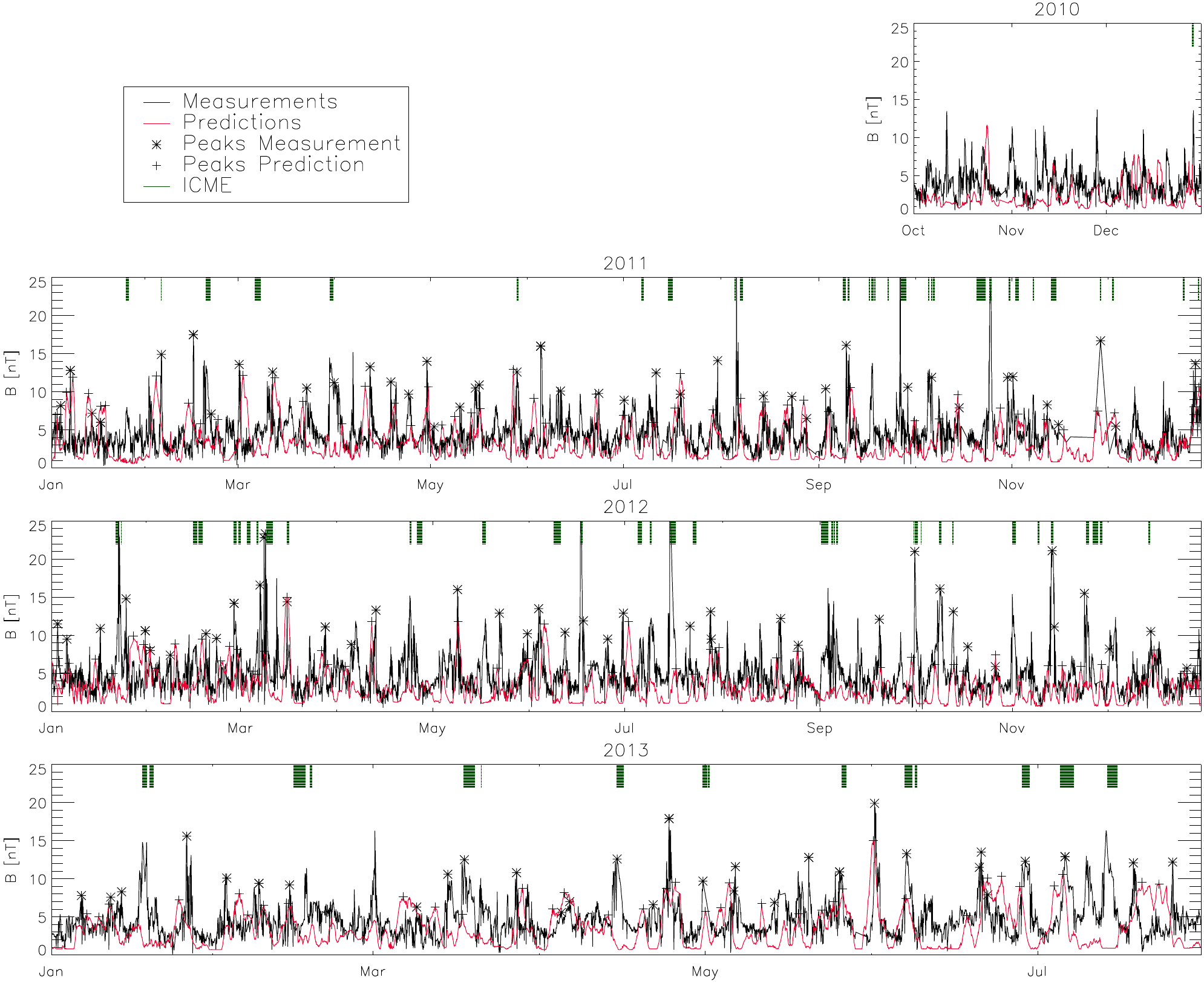}
      	\caption{Same as Fig.~\ref{fig007} but for the total magnetic field strength.}
				\label{fig008}
   \end{figure}

Figures \ref{fig007} and \ref{fig008} show the solar wind speed $v$ and magnetic field strength $B$ predictions (red lines) based on the adaptive algorithm together with \textit{in situ} measurements by ACE (black lines) at 1~AU for the years 2010-2013. The algorithm begins with the 3rd of October 2010, calculating the prediction function for CR2105 based on CR2102-2104. Cross-signs mark peaks in predicted solar wind speed and stars depict peaks in the corresponding measurements. Green bars depict the start and the end of identified ICMEs at 1 AU taken from Richardson and Cane list. The algorithm adapts to the variations of the slow solar wind during a year. The values of the predicted slow solar wind speed follow the trend in the measurements (see Fig.~\ref{fig007} and Fig.~\ref{fig008}). 

\begin{figure}[ht]    
					\includegraphics[width =\linewidth]{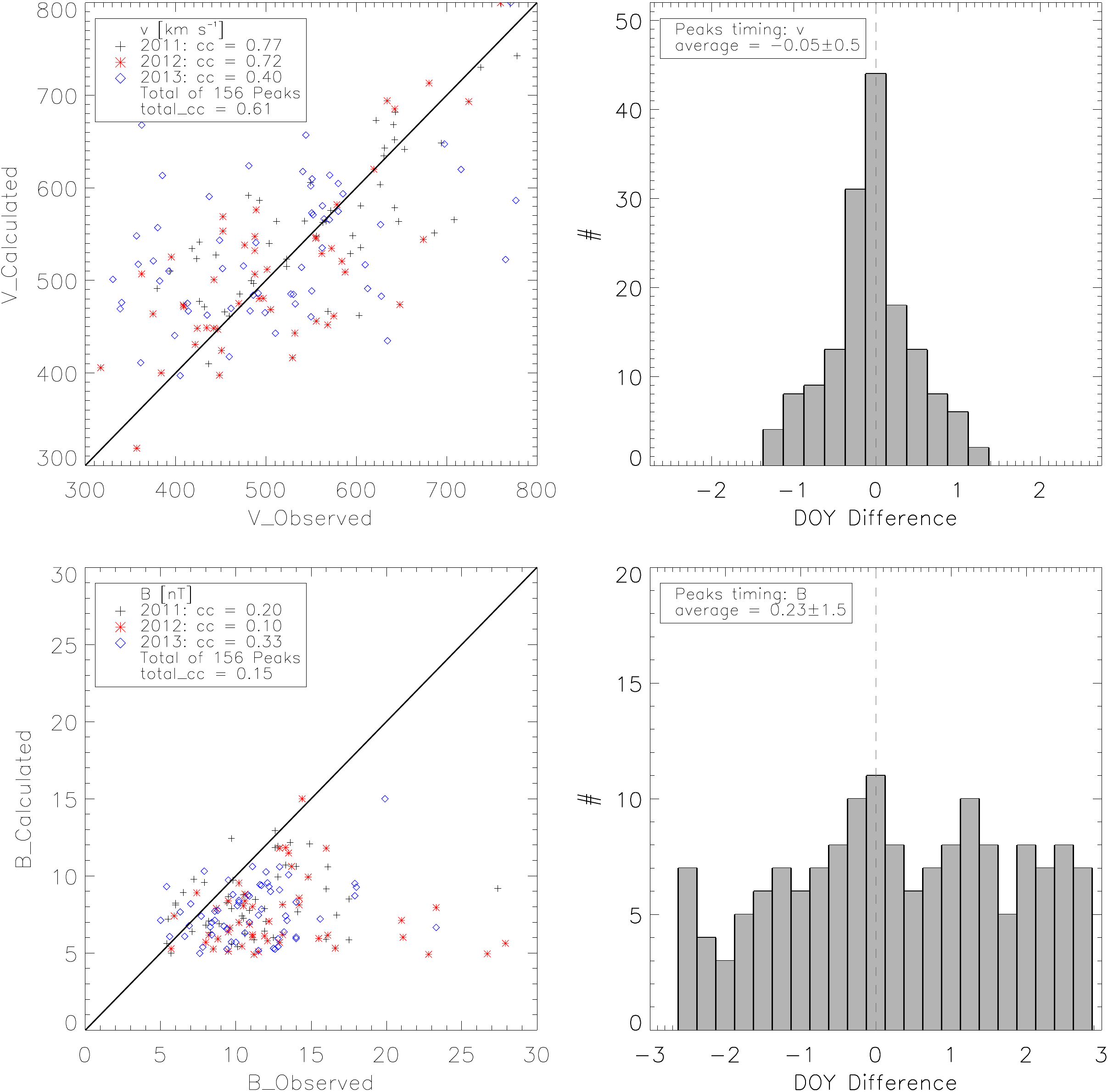}
      	\caption{Left panels: Scatter plot of predicted versus measured peaks in the solar wind speed (top) and total magnetic field strengths (bottom). In total, $156$ peaks were identified in the CH areas during the years 2011 - 2013. The related correlation coefficients are indicated in the insets (2011: plus sign; 2012: red stars; 2013: blue diamonds). The one-to-one correspondence is plotted as a black line. Right panels: Histograms of the difference of arrival times of measured and predicted solar wind $v$ and $B$.}
				\label{fig009}
   \end{figure}

Figure~\ref{fig009} shows scatter plots of 156 peaks identified in predicted and measured solar wind speed $v$, and magnetic field strength $B$ (indicated in Figures~\ref{fig007}-\ref{fig008}) for the years 2011-2013. The peaks were automatically determined by scanning for local maxima in the extracted fractional areas $A$ data that exceed a value of $0.15$. It automatically searches for the closest peak in the measured solar wind data in a given $\pm$3~days window.

We derived correlation coefficients $cc$ for every year separately for the peaks in $v$ as well as $B$. For the solar wind speed we derived $cc_{\, v\,2011} = 0.77$, $cc_{\, v\,2012} = 0.72$ and $cc_{\,v\,2013} = 0.40$. Correlating all measured versus predicted peaks we found $cc_{v\,total} = 0.61$. For the magnetic field strength we found lower correlations with $cc_{\, B\,2011} = 0.20$, $cc_{\, B\,2012} = 0.10$ and $cc_{\,B\,2013} = 0.33$ and a $cc_{B\,total} = 0.15$. 

The histograms in Fig.~\ref{fig009} depict the differences in the arrival time between measured and predicted peaks. The average prediction difference for $v$ lie at $-0.05\pm0.51$ days. 
The histogram reveals that for 80$\%$ of the HSS peaks identified, the predicted arrival times lie within $\pm0.5$ days of the observed peak arrivals. For $B$ we found an average of $0.23\pm1.53$ days, but the histogram is broadly distributed over $\pm3$ days.

\section{Discussion and Conclusion} 
      \label{S-Conclusion} 
The adaptive algorithm delivers reliable predictions for the peaks in the undisturbed solar wind speed (i.e. not affected by ICMEs) with a lead time of 4 days. The prediction accuracy of the HSS arrival times lies within $\pm0.5$ days in $80\%$ of the cases. The predicted amplitudes reveal a correlation of $cc = 0.6$ over the three years under study. Predictions for the solar wind speed peaks at 1 AU based on this empirical model are thus comparable to numerical models such as ENLIL and WSA (tested for the year 2007 by \opencite{Gressl2014})

\inlinecite{Gressl2014} performed a detailed study of ENLIL model runs with solar wind parameters at 1~AU for the year 2007. The numerical solar wind models produce the best results for the parameter solar wind speed. During times of low solar activity the numerical models predicted the arrival times of HSSs to have typical uncertainties of approximately 0.5-1.5 days with cross-correlation coefficients in the range of $0.5$ to $0.7$ (see also \opencite{Lee2009,Jian2011,Gressl2014}). Also, the outcome for the numerical models was found to be dependent on synoptic magnetic map used as an input as stated by \opencite{riley2012}. 

In the case of the solar wind magnetic field strength, the adaptive algorithm using linear prediction functions, does not work.  The algorithm underestimates the magnetic field strength. In our current algorithm we derive the magnetic field by observing the sun in EUV and excluding any further information.  It was shown that the primary driver of geomagnetic disturbances is the coupling of $B$ and $v$ (\opencite{Siscoe1974,Crooker1977}). $vB$ (related to the electric field in the current sheet), $vB^{2}$ (Poynting flux) determine the level of geomagnetic activity (\opencite{Perreault1978}). $v^{2}B^{2}$ describes the energy coupling function for the magnetospheric substorm and the delivered power (see also \opencite{Kan1979}). One has to consider that magnetospheric current systems may respond differently to the HSS related and the CME-related solar wind disturbances (cf. \opencite{Verbanac2011,Verbanac2013} and references therein). This behavior shows that the combination of the solar-wind speed and magnetic field is essential to the process of energy transfer from the solar wind to the magnetosphere.
 Furthermore, numerical models such as ENLIL and WSA underestimate the magnetic field strength by a factor of two and the performance of the prediction models is very sensitive to the input synoptic map, temperature and density values and the computational-grid resolution used (\opencite{Gressl2014,Stevens2012,riley2012}).
Also, CME have huge impact on the interplanetary background solar-wind structure. The effects of CMEs on the solar-wind conditions are observed by the in-situ measurements but do not appear in the predictions of the background solar wind.

\begin{figure}[ht]    
					\includegraphics[width =\linewidth]{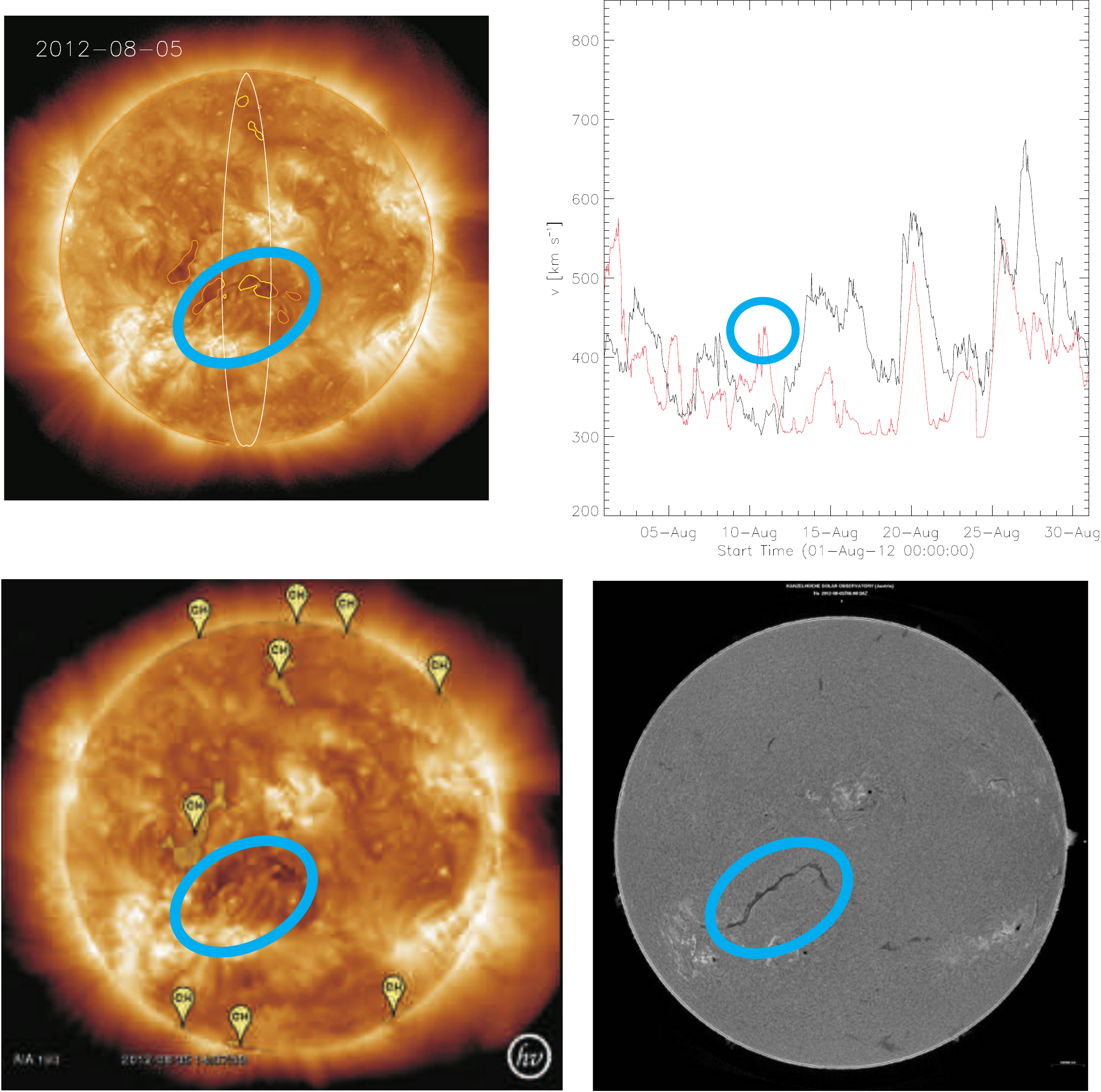}
      	\caption{Left top: Coronal holes detected by the algorithm for the SDO/AIA-193\AA\, image taken on 5th August, 2012. Right top: Measurements by ACE (black line) and predictions (red line) based on $A$. Bottom left: Coronal hole detections based on SPoCA-Suite. Bottom Right: KSO high contrast H$\alpha$ image.}
				\label{fig010}
   \end{figure}

Figure~\ref{fig010} illustrates that unreliable solar wind predictions can be caused by false detections of CHs in EUV images. We show coronal holes falsely detected by our algorithm (left top outlined with a blue circle) for the SDO/AIA-193\AA\, image taken on the 5th of August, 2012. In comparison, the \textit{Spatial Possibilistic Clustering Algorithm} SPoCA-Suite (\opencite{Delouille2012,delouille2014})  did not identify the filament as coronal hole (Image taken from helioviewer (\url{http://www.helioviewer.org/}). In the bottom right panel an H$\alpha$ image recorded at Kanzelhoehe observatory (\url{http://cesar.kso.ac.at/synoptic/ha_years.php}; \opencite{Poetzi2013}), clearly outlines the filament giving evidence of the current weakness of the algorithm to distinct between coronal holes and filaments, as they are both characterized by low intensities in the EUV images (see also \opencite{deToma2005}). We are currently developing a geometrical classification model that will be implemented in the algorithm to reduce coronal hole classification errors by taking advantage of filaments being asymmetric structures aligned along a preferred direction (\opencite{Reiss2014}).

\begin{figure}[ht]    
					\includegraphics[width=\linewidth]{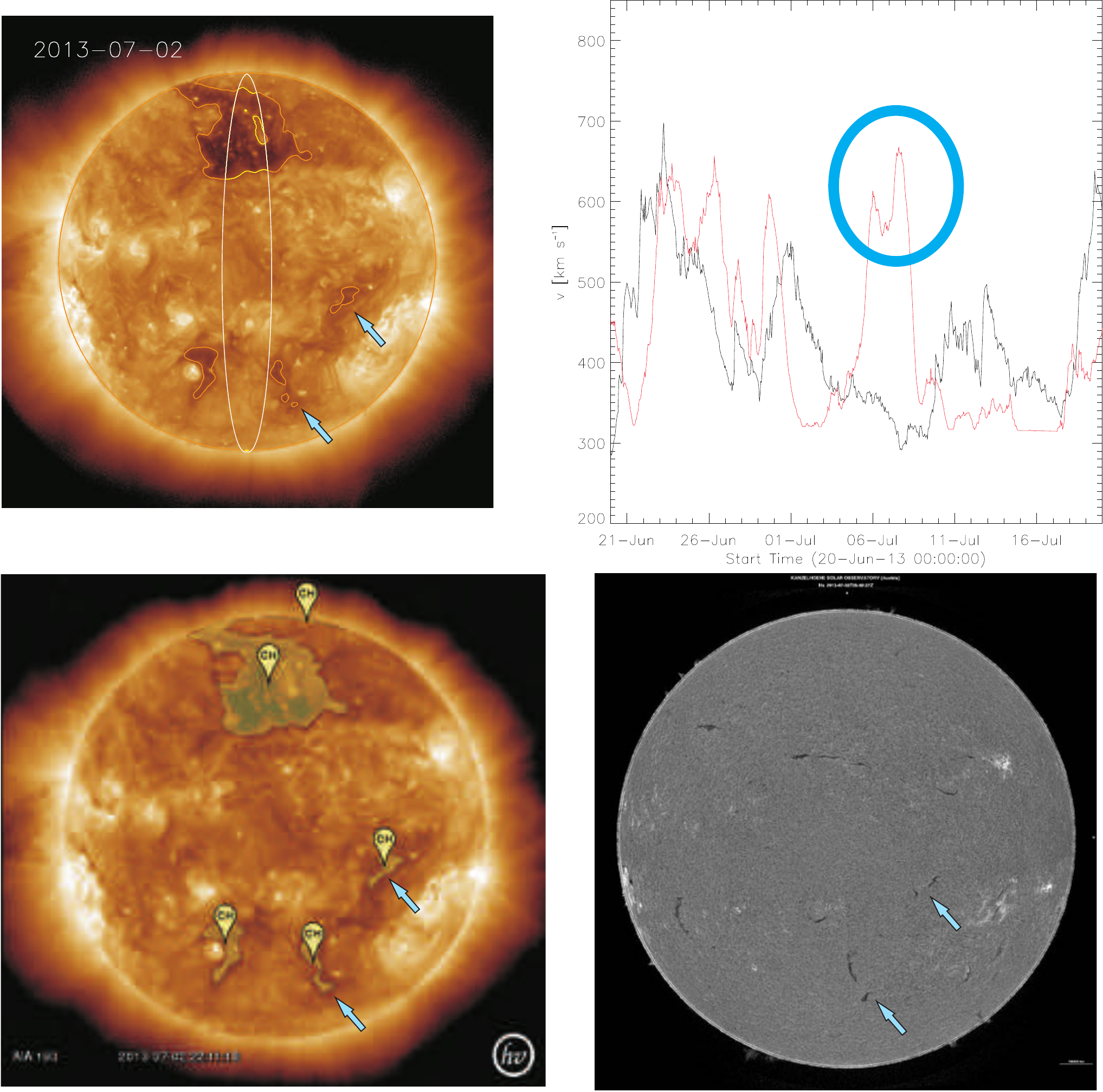}
      	\caption{Same as Figure~\ref{fig010} showing good agreement of the used algorithm (top left) and SPoCA-Suite (bottom right). But for 2nd July 2013 false solar wind speeds were predicted (black circle). Blue arrows point at small features wrongly detected by SPoCA and our algorithm.}
				\label{fig011}
   \end{figure}

\inlinecite{Schwenn2006} discusses that during low solar activity there appear no warps in the heliospheric current sheet (HCS) and the HCS lies close to the heliographic equator. So it can occur that we observe large polar holes and predict an increase in solar wind speed but the HSS is actually simply ``missing'' Earth. Figure~\ref{fig011} displays an example of a large coronal hole in the northern hemisphere on 2nd of July 2013. On this image, the SPoCA-Suite agrees with the detections of the coronal holes identified by our image recognition algorithm. The SPoCA-Suite also has the problem distinguishing between filaments and coronal holes, since it deals with intensity values.

 In general we note that in cases where no ICMEs are disturbing the solar wind parameters, the algorithm delivers reliable predictions of the solar wind speed at 1~AU from solar observations with an lead time of about 4~days. The algorithm is applied in real-time HSS forecasting, being hourly updated with the most recent SDO/AIA 193\AA \, image (\url{http://swe.uni-graz.at/solarwind}). It is a sturdy and fast image processing algorithm that can utilize EUV images in manner of seconds and works for several satellite missions and different instruments.

\begin{acks}

Courtesy of NASA/SDO and the AIA teams. We acknowledge the ACE SWEPAM and MAG instrument teams and the ACE Science Center. The research leading to these results has received funding from the European Commission Seventh Framework Programme (FP7/2007-2013) under the grant agreement FP7 No. 263252 (COMESEP). M.T. acknowledges the Austrian Science Fund (FWF): V195-N16. T.R. gratefully acknowledge support from NAWI Graz and the Forschungsstipendium by the University of Graz.

\end{acks}

\bibliographystyle{spr-mp-sola}
\tracingmacros=2

\end{article} 
\end{document}